\documentclass[letterpaper]{article} 
\usepackage{aaai2026}  
\usepackage{times}  
\usepackage{helvet}  
\usepackage{courier}  
\usepackage[hyphens]{url}  
\usepackage{graphicx} 
\usepackage{natbib}  
\usepackage{caption} 
\usepackage{algorithm}
\usepackage{algorithmic}

\usepackage{newfloat}
\usepackage{listings}
\DeclareCaptionStyle{ruled}{labelfont=normalfont,labelsep=colon,strut=off} 
\floatstyle{ruled}
\newfloat{listing}{tb}{lst}{}
\floatname{listing}{Listing}

\usepackage{amssymb}
\usepackage{amsmath}
\usepackage{tabularx}
\usepackage{enumitem}
\usepackage{subfigure}
\usepackage{comment}
\usepackage{xcolor}
\title{Dual-Head Physics-Informed Graph Decision Transformer for Distribution System Restoration}
\author {
    Hong Zhao\textsuperscript{\rm 1},
    Jin Wei-Kocsis\textsuperscript{\rm 1}\thanks{Corresponding author.},
    Adel Heidari Akhijahani\textsuperscript{\rm 2},
    Karen L Butler-Purry\textsuperscript{\rm 2},
}
\affiliations {
    \textsuperscript{\rm 1}Purdue University, West Lafayette\\
    \textsuperscript{\rm 2}Texas A\&M University, College Station\\
    \{zhao1211,kocsis0\}@purdue.edu, adelheidari@exchange.tamu.edu, klbutler@tamu.edu
}

\usepackage{bibentry}

\begin{document}

\maketitle
\begin{abstract}
Driven by recent advances in sensing and computing, deep reinforcement learning (DRL) technologies have shown great potential for addressing distribution system restoration (DSR) under uncertainty. However, their data-intensive nature and reliance on the Markov Decision Process (MDP) assumption limit their ability to handle scenarios that require long-term temporal dependencies or few-shot and zero-shot decision making. Emerging Decision Transformers (DTs), which leverage causal transformers for sequence modeling in DRL tasks, offer a promising alternative. However, their reliance on return-to-go (RTG) cloning and limited generalization capacity restricts their effectiveness in dynamic power system environments. To address these challenges, we introduce an innovative Dual-Head Physics-informed Graph Decision Transformer (DH-PGDT) that integrates physical modeling, structural reasoning, and subgoal-based guidance to enable scalable and robust DSR even in zero-shot or few-shot scenarios. DH-PGDT features a dual-head physics-informed causal transformer architecture comprising Guidance Head, which generates subgoal representations, and Action Head, which uses these subgoals to generate actions independently of RTG. It also incorporates an operational constraint-aware graph reasoning module that encodes power system topology and operational constraints to generate a confidence-weighted action vector for refining DT trajectories. This design effectively improves generalization and enables robust adaptation to unseen scenarios. While this work focuses on DSR, the underlying computing model of the proposed PGDT is broadly applicable to sequential decision making across various power system operations and other complex engineering domains.
\end{abstract}

\section{Introduction}

Recent extreme events, such as the massive power outage in Houston caused by Hurricane Beryl, have underscored the urgent need to enhance the resilience of modern power systems. The distribution system plays a pivotal role in delivering electricity to end-users, and its ability to recover swiftly from disruptions is crucial to overall grid resilience. A key indicator of distribution system resiliency is the ability to restore service to critical loads following disruptions, known as Distribution System Restoration (DSR)~\cite{Chen2017ModernizingDS}. Forming microgrids (MGs) with dynamic boundaries as a service restoration strategy is a promising solution for enabling effective DSR~\cite{9693132,9765477,9508140}. By incorporating various energy sources, distributed generators (DGs), along with remotely controlled switches, a distribution system can be partitioned into multiple self-sufficient MGs. This strategy enables rapid reconfiguration and localized recovery, significantly enhancing system adaptability and restoration efficiency. In this paper, we focus on advancing DSR through the sequential formation of MGs with dynamic boundaries to improve distribution system resiliency under uncertainty. 

Traditional methods for DSR, such as mathematical programming, heuristic algorithms, and expert systems, often suffer from limited scalability and robustness. In recent years, machine learning, particularly deep reinforcement learning (DRL), has emerged as a powerful approach for enabling more efficient, adaptive, and robust DSR~\cite{WU2019122487, 9282132, 9126832, 9705112}. However, DRL methods typically rely on Markov Decision Process (MDP) or partially observable MDP assumptions and require extensive training data, which limits their applicability to large-scale DSR scenarios that demand long-horizon temporal reasoning or operate under few-shot or zero-shot conditions.

The emergence of Decision Transformers (DTs), which adapt causal transformer architectures for scalable sequential decision-making while effectively capturing long-term temporal dependencies, has introduced a compelling alternative to conventional DRL methods~\cite{ChenLuRajeswaran21}. However, their application to DSR remains largely unexplored. To the best of our knowledge, our prior work~\cite{ZhaoKocsisAkhijahani25} is the first to investigate the use of DTs in DSR operations. Despite their potential, their reliance on RTG cloning and limited generalization capacity can restrict their effectiveness in dynamic power system environments. Recent studies have attempted to address the generalization limitation of DTs from various perspectives. A goal-conditioned DT with multi-objective pretraining was proposed using explicit goals and time-to-goal inputs~\cite{FuLongChen24}. Although effective in surgical robotics, its core idea on goal variability limits applicability to DSR, where the restoration objective remains consistent. A conservative DT was proposed to avoid RTG at inference by using a separate transformer to estimate rewards conservatively~\cite{KimNohJang23}. However, the method still assumes MDP and significantly increases computational cost due to the use of dual transformers. In~\cite{MaXiaoLiang23}, the generalization limitation was addressed by introducing a hierarchical reformulation with auto-tuned high-level prompts and joint optimization, and, in~\cite{HsuBozkurtDong24}, the limitation was addressed by replacing RTG with a temporal-difference-learned steering signal. Both methods retain MDP assumptions. Trajectory prompt-based approaches were introduced in~\cite{XuShenZhang22,YaoChenYao25}, which condition DTs on short trajectory demonstrations for few-shot generalization. Although promising, their reliance on high-quality demonstrations limits applicability to DSR in unforeseen scenarios. Therefore, while effective in their respective application scenarios, these DT variants are not well suited for DSR, motivating the development of our proposed framework, which is explicitly designed to overcome these limitations in DSR settings characterized by uncertainty, complex constraints, and limited demonstration data. 

In this paper, we propose a novel Dual-Head Physics-informed Graph Decision Transformer (DH-PGDT) that enables robust and scalable DSR even in unforeseen operation scenarios with complex physical and operational constraints. DH-PGDT integrates an operational constraint-aware graph reasoning module with a dual-head physics-informed causal transformer architecture to incorporate physical and operational information while reducing reliance on RTG. The main contributions of our proposed work are twofold: (1) developing a novel DH-PGDT framework that enables physical and operational constraint-aware and robust DSR decisions under uncertain and unforeseen conditions; and (2) conducting comprehensive evaluations demonstrating the effectiveness of DH-PGDT in terms of restoration efficiency, constraint satisfaction, and zero-shot/few-shot adaptability. Additionally, although our current work focuses on DSR, the underlying computing framework is broadly applicable to sequential decision-making tasks in other complex engineering domains.

The next section illustrates the problem settings for our work. After this, we describe our proposed DH-PGDT framework for the DSR decision making. The following section shows the case studies and performance evaluations of the proposed DH-PGDT framework. Conclusions are presented in the last section.

\section{Problem Settings}\label{section:Problem_Settings}
\subsection{DSR Problem Modeling}
In the early stage of this research, the proposed DSR method is modeled as a sequential decision-making process involving sequences of control actions on switches to restore loads to their normal operational states. These sequences of switching actions, referred to as energization branches, are each associated with a single energy source (e.g., a single active distributed generator (DG)). Specifically, each energization branch begins at the switch connected to the associated energy source and aims to maximize load restoration by forming multiple microgrids, while ensuring compliance with all operational and physical constraints. Additionally, to reduce the space of control actions in the modeling, we adopt the concept of node cell as introduced in \citealp{8587147}. A node cell is defined as a set of nodes that are interconnected directly by non-switchable lines. Consequently, all the lines and loads within a node cell will be energized simultaneously. Furthermore, in our current work, the physical constraints include: 1) the node voltage limits, 2) power flow constraints, 3) DGs’ generation capacities, and 4) DGs' prime mover ability. The operational constraints include 1) prevention of loop formation and 2) ensuring no node cell is visited more than once. 

\subsection{DSR Problem Formulation within a DT Framework}
We further formulate the sequential decision-making process for energization branches in the DSR problem model. Table~\ref{Definitions} shows the definitions of the parameters and variables used in this paper.
\begin{table}[!h]
    \centering
    \begin{tabular}{>{\centering\arraybackslash}m{2cm}|>{\centering\arraybackslash}m{5cm}}
\hline
$s_t, \mathcal{S}$ & State at time $t$, the state space \\ \hline
$a_t, \mathcal{A}$ & Action at time t, the action space \\ \hline
$R$ & Reward \\ \hline
$\mathcal{P}$ & Transition probability \\ \hline
$r_t$ & Reward at time t \\ \hline
$T$ & Time Horizon \\ \hline
$L$ & Set of all the loads \\ \hline
$N$ & Set of all buses \\ \hline
$C, |C|$ & Set of all node cells, number of node cells \\ \hline
$E$ & Set of all DGs \\ \hline
$x_{i,t}^L, x_{i, t}^C$ & Energization status of Load $i$ at time $t$, energization status of node cell $i$ at time $t$ \\ \hline
$P_{l,t}^L, P_{l,t}^C, P_{l, t}$ & Active power of Load $l$ at time $t$ when restored, accumulative active power of all loads in node cell $l$ at time $t$ when restored, nominal active power of Load $l$ at time $t$. \\ \hline
$P_{dg,t}^E$ & Active output power of DG $dg$ at time $t$. \\ \hline
$P_{dg}^{\text{ramp}}$ & Maximum absolute change of active power settings of DG $dg$ \\ \hline
$H_{i,t}^L$ & Squared voltage magnitude of Load $i$ at time $t$ \\ \hline
$H_i^{\min},H_i^{\max}$ & Minimum and maximum squared nodal voltage of Load $i$ \\ \hline
$V_{l,t}^{L,p}$ & Voltage penalty function of Load $l$ at time $t$ \\ \hline
$S_{dg,t}^{E,p}$ & Ramp rate penalty function of DG $dg$ at time $t$ \\ \hline
$\theta$ & Learnable parameters for the proposed DH-PGDT \\ \hline
$q$ & Number of subgoals for the DH-PGDT \\ \hline
\end{tabular}
    \caption{Definition of variables and parameters}
    \label{Definitions}
\end{table}

In the context of DT, the key components in the DSR problem are formulated as follows:
\begin{itemize}
    \item State $\mathbf{s}_{t}\in\mathcal{S}$ is defined as the available observation vector of the overall distribution system at time $t$. The state vector $\mathbf{s}_t = \left[ \mathbf{s}_{1,t} \,\|\, \mathbf{s}_{2,t} \right]$ is defined as the concatenation of two vectors, $\mathbf{s}_{1,t}$ and $\mathbf{s}_{2,t}$:
    \begin{itemize}
        \item $\mathbf{s}_{1,t}$ is a binary vector representing energization status of individual node cells. If node cell $j$ is chosen to be energized at step $t$, then the $j$-th element will always be $1$ at any time step larger than $t$.
        \item $\mathbf{s}_{2,t}$ is another binary vector representing whether the individual node cells are activated at the current time step $t$. Number of $1$'s in $\mathbf{s}_{2,t}$ will always equal the number of energization branches.
    \end{itemize}
    \item Action $\mathbf{a}_t \in \mathcal{A}$ is defined as a vector of confidence probabilities over the operable switches, where each element represents the model’s confidence in activating a specific switch. During DSR, switches are sampled from $a_t$ to determine which to activate, aiming to maximize load restoration while satisfying operational and physical constraints.
    \item Time horizon $T$ defines the total number of decision-making steps.
    \item Physics-informed Return-to-go (RTG) $\hat{R}_t$ is defined as $\hat{R}_t=\sum_{j=t}^Tr_j$, which is the sum of rewards $r_t$ from a given time to the end of the time horizon. The rewards $r_t$ evaluate the effectiveness of action $\mathbf{a}_t$ taken in state $\mathbf{s}_t$ in achieving the objective of our DSR problem, which is to maximize the load restoration in the time horizon $T$ while adhering to physical constraints. Therefore, the reward will be formulated as:
    \begin{equation}
        \label{Reward Formation}
        \begin{split}
            r_t=R_t^A(\mathbf{s}_t, \mathbf{a}_t)+w_{p_1}\times R_t^V(\mathbf{s}_t, \mathbf{a}_t)+w_{p_2}\times R_t^E(\mathbf{s}_t, \mathbf{a}_t)
        \end{split}
    \end{equation}
    where:
    \begin{itemize}
        \item $R_t^A(\mathbf{s}_t,\mathbf{a}_t)$ is the reward related to the total active power restoration at $t$, which is defined as:
    \begin{equation}
        \label{Restoration reward}
        R_t^A (\mathbf{s}_t,\mathbf{a}_t )=\left(\sum_{l\in L}x_{l,t}^L P_{l,t}^L\right)\times\Delta t
    \end{equation}
    By incorporating the concept of node cell, Eq.~(\ref{Restoration reward}) can be rewritten as:
    \begin{equation}
        \label{Restoration reward node cell}
        R_t^A (\mathbf{s}_t,\mathbf{a}_t )=\left(\sum_{l\in C}x_{l,t}^CP_{l,t}^C\right)\times\Delta t
    \end{equation}
    \item $R_t^V(\mathbf{s}_t,\mathbf{a}_t)$ is a penalty term to penalize actions that violate the voltage constraints. It is defined as:
    \begin{equation}
        \label{Penalty for voltage}
        R_t^V(\mathbf{s}_t,\mathbf{a}_t)=-\left(\sum_{l\in L}x_{l,t}^LV_{l,t}^{L,p}\right)\times\Delta t
    \end{equation}
    where
    \begin{equation}
        \label{Individual penalty for voltage}
        V_{l,t}^{L,p}=\max\left(0, H_{l,t}^L-H_l^{\max}\right)+\max\left(0, H_l^{\min}-H_{l,t}^L\right)
    \end{equation}
    Where $V_{l,t}^{L,p}$ represents the penalty terms associated with individual loads to ensure their node voltage magnitudes do not violate the constraints. These constraints are formulated using squared nodal voltage range $[H_i^{\min},H_i^{\max}]$.
    \item $R_t^E(\mathbf{s}_t,\mathbf{a}_t)$ is a penalty term to penalize actions that violate the DG ramp rate constraints. It is defined as:
    \begin{equation}
        \label{Penalty for DG ramp rate}
        R_t^E(\mathbf{s}_t,\mathbf{a}_t)=-\left(\sum_{dg\in E}S_{dg,t}^{E,p}\right)\times\Delta t
    \end{equation}
    where
    \begin{equation}
        \label{Individual penalty for DG ramp rate}
        \begin{split}
        S_{dg,t}^{E,p}&=\max\left(0, (P_{dg,t}^E-P_{dg,t-1}^E)-P_{dg}^{\text{ramp}}\right)\\
        &+\max\left(0, -P_{dg}^{\text{ramp}}-(P_{dg,t}^E-P_{dg,t-1}^E)\right)
        \end{split}
    \end{equation}
    Where $S_{dg,t}^{E,p}$ represents the penalty terms associated with individual DGs to ensure their absolute change in the active power output do not exceed given maximum values. Alternatively, this is to say that the change in active power output $P_{dg,t}^E-P_{dg,t-1}^E$ for DG $dg$ at time $t$ lies in the range $[-P_{dg}^{\text{ramp}},P_{dg}^{\text{ramp}}]$.
    \end{itemize}
    The weight terms $w_{p_1}$ and $w_{p_2}$ are set to ensure that the penalties are comparable to the active restored power term. 
    \item Operational constraint-aware mask function $f\left(\mathbf{s}_t,\cdot\right)$ is formulated to automate the enforcement of operational constraints. Its output is a mask vector $\mathbf{M}$ that is used to filter out actions that violate these constraints during decision making.
\end{itemize}

\section{Proposed DH-PGDT Framework for DSR Decision
Makings}

\begin{figure}[t]
\centering
\includegraphics[width=0.9\columnwidth]{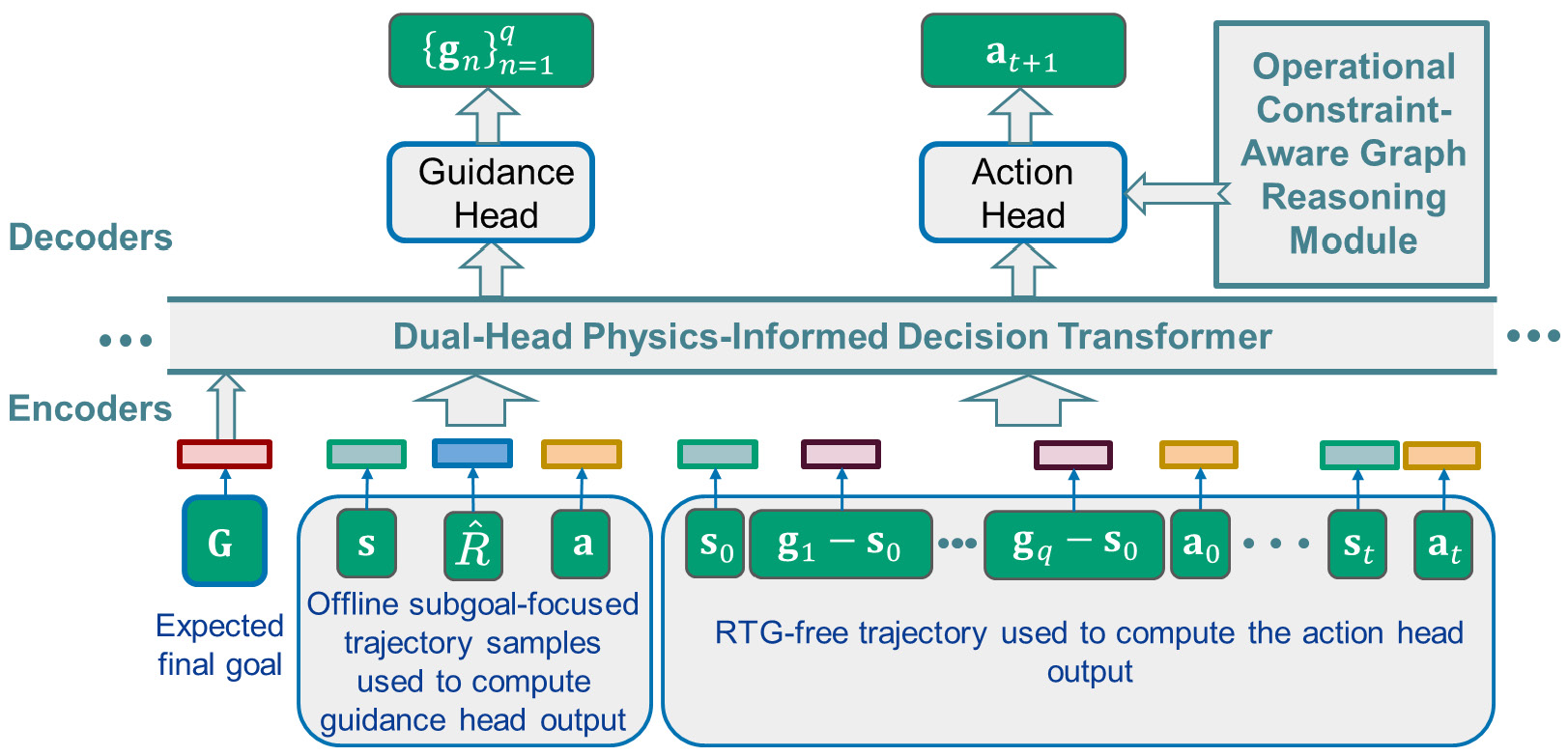}
\caption{Overview of the architecture of our proposed DH-PGDT framework.}
\label{fig:DH-PGDT overview}
\end{figure}

In this section, we will introduce our proposed DH-PGDT. As illustrated in Fig.~\ref{fig:DH-PGDT overview}, the DH-PGDT mainly consists of a dual-head physics-informed causal transformer architecture and an operational constraint-aware graph reasoning module.

\subsection{Dual-Head Physics-Informed Causal Transformer Architecture}
As shown in Fig.~\ref{fig:DH-PGDT overview}, our dual-head physics-informed causal transformer architecture mainly comprises: 1) an encoder that consists of linear layers followed by an embedding layer; 2) a GPT-based dual-head physics-informed causal transformer with a causal self-attention mask; and 3) a decoder. The encoder processes offline subgoal-focused or RTG-free trajectory samples into token embeddings, which are input to the GPT-based dual-head physics-informed causal transformer to compute the outputs of the guidance and action heads. The action head output is then passed to the decoder, which maps the predicted trajectory embeddings back to the original action space, producing the final control actions for the operational switches in DSR tasks. 

The GPT-based dual-head physics-informed causal transformer lies at the core of our proposed architecture. It integrates Guidance Head and Action Head to achieve both effectiveness and robustness in decision-making for DSR. In the following, we present the detailed design of the dual-head transformer.

\subsubsection{Guidance Head Design:} 

The \textit{Guidance Head} is designed to predict a sequence of subgoal representations \( \{\mathbf{g}_n\}_{n=1}^{q} \) in the state space \( \mathcal{S} \), conditioned on the expected final goal \( \mathbf{G} \) and a subgoal-focused trajectory \( \tau^g \), as illustrated in Fig.~\ref{fig:DH-PGDT overview}. Both the final goal \( \mathbf{G} \) and the predicted subgoals \( g_n \) lie in the same state space \( \mathcal{S} \). In our current early-stage implementation, the final goal \( \mathbf{G} \) is defined as the fully energized target state:
\[
\mathbf{G} = \mathbf{s}_T = \left[ \mathbf{s}_{1,T} \,\|\, \mathbf{s}_{2,T} \right],
\]
where \( \mathbf{s}_{1,T} \) is a vector of all ones (indicating that all node cells are energized), and \( \mathbf{s}_{2,T} \) is masked. To construct subgoals, in our current work, we predefine a fixed number \( q \) of intermediate subgoals. Each subgoal \( g_n \) corresponds to the state \( \mathbf{s}_{m_n} \) when $\lceil nN/(q+1) \rceil$ node cells are energized, where \( N \) is the total number of node cells to be restored. Thus, the subgoals represent meaningful intermediate milestones toward the final restoration goal $\mathbf{G}$. If a trajectory terminates before reaching certain subgoal thresholds (i.e., fewer than \( \lceil sN/(q+1) \rceil \) nodes energized for some \( s \leq q \)), we assign those remaining subgoals to the final timestep \( T \) for simplicity. The trajectory input to the Guidance Head, \( \tau^g \), is a \textit{subgoal-focused trajectory} sparsely sampled from the dataset of offline trajectories. It only includes the initial state $\mathbf{s}_0$, action $\mathbf{a}_0$, and physics-informed RTG $\hat{R}_0$ and the subset of transitions associated with the subgoal steps. The RTG values \( \hat{R} \) is described in the previous section. Formally, the trajectory is defined as:
\begin{equation}
\label{Adjusted Input Sequence for Guidance Generation}
\tau^g = \left\{ \mathbf{G}, \mathbf{s}_0, \mathbf{a}_0, \hat{R}_0, \tau_s^g \right\},
\end{equation}
where the subgoal-related sequence \( \tau_s^g \) is defined as:
\begin{equation}
\tau_s^g = \left\{ \mathbf{s}_{t_{m_1}}, \mathbf{a}_{t_{m_1}}, \hat{R}_{t_{m_1}}, \dots, \mathbf{s}_{t_{m_{q}}}, \mathbf{a}_{t_{m_{q}}}, \hat{R}_{t_{m_{q}}} \right\},
\end{equation}
and \( \hat{R} \) is the physics-informed RTG defined in the previous section.

\subsubsection{Action Head Design:} 

Guided by the subgoal representations generated from the Guidance Head, the Action Head is designed to predict control actions at each step independently of individual-action-level RTG values. This decoupling enhances robustness by avoiding RTG overfitting and enables more flexible trajectory stitching.

Suppose we have a state-action-RTG trajectory of length \( K \), denoted as:
\[
\tau_{-K:(K-1)} = \big\{\mathbf{s}_0, \mathbf{a}_0, \hat{R}_0, \dots, \mathbf{s}_{K-1}, \mathbf{a}_{K-1}, \hat{R}_{K-1}\big\}.
\] To generate inputs for the Action Head, we augment this trajectory by appending the final goal \( \mathbf{G} \) and the subgoal offsets \( \mathbf{g}_i - \mathbf{s}_0 \) to the beginning of the sequence. The full input sequence used for training is:
\begin{equation}
\label{Adjusted Input Sequence for Action Generation}
\begin{split}
\tau'_{-K:(K-1);q} = \big\{&\mathbf{G}, \mathbf{s}_0, \mathbf{g}_1 - \mathbf{s}_0, \mathbf{g}_2 - \mathbf{s}_0, \dots, \mathbf{g}_{q} - \mathbf{s}_0, \mathbf{a}_0, \\
&\mathbf{s}_1, \mathbf{a}_1, \dots, \mathbf{s}_{K-1}, \mathbf{a}_{K-1} \big\},
\end{split}
\end{equation}
where \( q \) is the predefined number of subgoals.

For generality and ease of notation, we also define a time-shifted sequence:
\begin{equation}
\begin{split}
\tau'_{-K:t;q} = \big\{&\mathbf{G}, \mathbf{s}_{t-K+1}, \mathbf{g}_1 - \mathbf{s}_{t-K+1}, \mathbf{g}_2 - \mathbf{s}_{t-K+1}, \dots, \\
&\mathbf{g}_{q} - \mathbf{s}_{t-K+1}, \mathbf{a}_{t-K+1}, \\
&\mathbf{s}_{t-K+2}, \mathbf{a}_{t-K+2}, \dots, \mathbf{s}_t, \mathbf{a}_t \big\},
\end{split}
\end{equation} which represents the same sequence structure aligned with a later time index \( t \). This representation \( \tau'_{-K:t;q} \) will be used in the subsequent sections describing the model training and testing pipelines.

\subsection{Operational Constraint-Aware Graph Reasoning Module}
To enforce operational constraints during action selection, we introduce an operational constraint-aware graph reasoning module integrated into the action head. This module automates operational constraint enforcement by leveraging real-time topological and operational information from the distribution system.
Specifically, we formulate an adjacency matrix $\mathbf{A}_t$ to represent the system topology at time step $t$, capturing the connectivity between energized node cells and candidate nodes that may be energized. The previously defined state vector $\mathbf{s}_t$, which encodes dynamic attributes for each node cell, such as energization status and whether it has been activated at the current step, is treated as a node feature matrix in the graph module.
Using $\mathbf{A}_t$ and $\mathbf{s}_t$, we define an operational constraint-aware mask function $f\left(\mathbf{s}_t,\mathbf{A}_t\right)$ that outputs a mask vector $\mathbf{M}_t$. This mask is applied to the raw logit vector $\tilde{a}_{t+1}$ generated by the action head, effectively filtering out infeasible actions, such as those that would create network loops or revisit previously energized nodes. The masked logits are then passed through a softmax layer to produce the final confidence probability vector, which constitutes the action head’s output $a_{t+1}$.

\subsection{Model Training and Inference Pipeline}

\subsubsection{Model Training:}

The overall model training procedure is described as follows:
\begin{enumerate}[label=\arabic*)]
    \item We prepare a dataset \( \mathcal{D} \) consisting of offline trajectories for DSR operations on the targeted distribution system. These trajectories can be collected from domain experts or generated via simple offline random walks that apply sequences of control actions to the operable switches.

    \item Once \( \mathcal{D} \) contains a sufficient number of offline trajectories, we preprocess each trajectory through two main steps. First, for each trajectory \( \tau \in \mathcal{D} \), we extract a subgoal-focused trajectory \( \tau_s^g \) as described in the Guidance Head Design. Second, using the expected final goal \( \mathbf{G} \), we construct a modified trajectory \( \tau_m \) as:
    \begin{equation}
    \label{Trajectories of New Dataset}
    \tau_m = \left(U_1, U_2, \dots, U_T\right),
    \end{equation}
    where each element is defined as
    \[
    U_t = (\mathbf{G}, \mathbf{s}_t, \mathbf{a}_t, \hat{R}_t, \tau_s^g), \quad t = 1, 2, \dots, T.
    \]
    The resulting modified trajectories are combined into a new training dataset \( \mathcal{D}' = \{ \tau_m \} \).

    \item Minibatches \( B \) of length \( K \) are sampled from \( \mathcal{D}' \) and fed into the DH-PGDT framework to compute control actions for DSR. The model parameters are updated by minimizing the joint loss:
    \[
    \mathcal{L} = \mathcal{L}_{g} + \mathcal{L}_{a},
    \]
    where \( \mathcal{L}_{g} \) is defined as the mean squared error between the predicted subgoal representations and the ground-truth subgoals, and \( \mathcal{L}_{a} \) is defined as the cross-entropy loss between the predicted and true control actions.

    \item Repeat Step 3 for \( M \) training episodes.
\end{enumerate}

The above procedure is summarized in Algorithm~\ref{DH-PGDT training}.

\begin{algorithm}[h]
\caption{DH-PGDT Model Training}
\label{DH-PGDT training}
\textbf{Initialize:} Offline dataset \( \mathcal{D} \), predefined goal \( \mathbf{G} \), DH-PGDT model with parameters \( \theta \), maximum number of episodes \( M \), and minibatch size \( b \).
\begin{algorithmic}[1]
\STATE Construct the processed dataset \( \mathcal{D}' \) from \( \mathcal{D} \) and \( \mathbf{G} \) using Eq.~\eqref{Trajectories of New Dataset}.
\FOR{episode \( m = 1 \) to \( M \)}
    \STATE Sample a random minibatch \( B \) of \( b \) sequences of length \( K \) from \( \mathcal{D}' \).
    \STATE Compute action and guidance predictions using DH-PGDT on batch \( B \).
    \STATE Calculate the loss \( \mathcal{L} = \mathcal{L}_g + \mathcal{L}_a \).
    \STATE Update model parameters \( \theta \).
\ENDFOR
\end{algorithmic}
\end{algorithm}

\subsubsection{Model Inference}
The inference process of our DH-PGDT model consists of two stages. First, given the initial state of the distribution system \( \mathbf{s}_0 \), the predefined final goal \( \mathbf{G} \), and the initial physics-informed RTG \( \hat{R}_0 \), the Guidance Head of DH-PGDT generates a sequence of subgoal representations \( \mathbf{g}_1, \mathbf{g}_2, \dots, \mathbf{g}_q \). For brevity, we denote this computation as \( \text{GH\_PGDT}(\cdot) \), referring to the Guidance Head module of the DH-PGDT model. Next, these subgoals are fed into the same DH-PGDT model along with the system states, actions, and the final goal \( \mathbf{G} \) to generate the next control action. We denote this computation as \( \text{AH\_PGDT}(\cdot) \), representing the Action Head module of DH-PGDT. The overall inference procedure is outlined in Algorithm~\ref{DH-PGDT inference}.\begin{algorithm}[h]
\caption{Inference of the Trained DH-PGDT Model}
\label{DH-PGDT inference}
\textbf{Initialize:} DH-PGDT model with trained parameters \( \theta \), predefined goal \( \mathbf{G} \), initial RTG \( \hat{R}_0 \), and initial state \( \mathbf{s}_0 \).
\begin{algorithmic}[1]
\STATE Set \( \mathbf{s}_1 \leftarrow \mathbf{s}_0 \), \( \tau_1 \leftarrow (\mathbf{s}_1, \hat{R}_0) \)
\STATE Obtain \( \mathbf{g}_1, \dots, \mathbf{g}_q \leftarrow \text{GH\_PGDT}(\tau_1) \)
\STATE Set \( \tau_2 \leftarrow \{\mathbf{G}, \mathbf{s}_1, \mathbf{g}_1 - \mathbf{s}_1, \dots, \mathbf{g}_q - \mathbf{s}_1\} \)
\FOR{time step \( t = 1 \) to \( T-1 \)}
    \STATE Compute action: \( \mathbf{a}_t \leftarrow \text{AH\_PGDT}(\tau_2) \)
    \STATE Apply action \( \mathbf{a}_t \) and observe next state \( \mathbf{s}_{t+1} \)
    \STATE Append \( (\mathbf{a}_t, \mathbf{s}_{t+1}) \) to \( \tau_2 \)
    \STATE Keep the last $K$ time steps of $\tau_2$ (i.e., $\tau_2\leftarrow\tau'_{-K:t+1;q}$).
\ENDFOR
\end{algorithmic}
\end{algorithm}

\section{Case Studies}
In this section, we evaluate the performance of our proposed DH-PGDT framework using three distribution system testbeds simulated in the Open Distribution Simulator Software (OpenDSS): a modified IEEE 13-node test feeder~\cite{IEEEStandard}, a modified IEEE 123-node test feeder~\cite{IEEEStandard}, and a modified Iowa 240-node test feeder~\cite{Bu2019ATD}. To evaluate the performance of our DH-PGDT framework, we compare it with a DT framework, PIDT, developed in our previous work~\cite{ZhaoKocsisAkhijahani25}, and two benchmark DRLs, the Proximal Policy Optimization (PPO) algorithm~\cite{Schulman2017ProximalPO}, and the Advantage Actor-Critic (A2C) algorithm \cite{Mnih2016AsynchronousMF}, in the DSR tasks.

\subsection{Modified IEEE 13-Node Test Feeder}
\begin{figure}[t]
\centering
\includegraphics[width=\columnwidth]{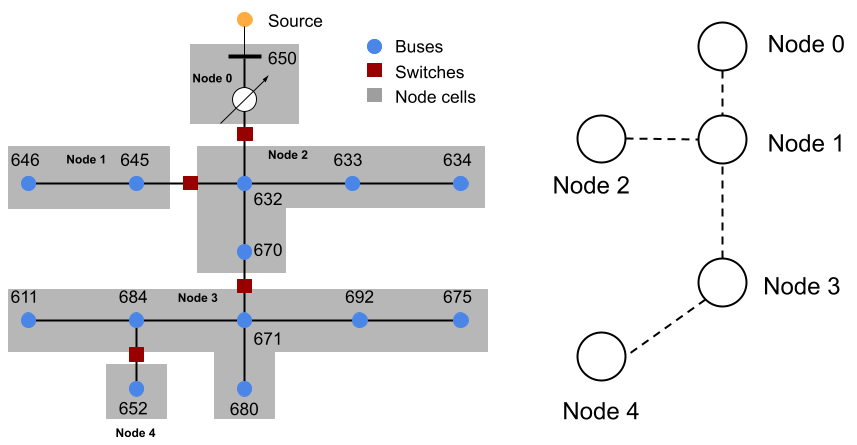}\\
~~~~~(a)~~~~~~~~~~~~~~~~~~~~~~~~~~~~~~~~~~~~~~~(b)
\caption{(a): The physical topology of the modified IEEE 13-node test feeder; and (b): The node-cell-based graph representation of the modified IEEE 13-node test feeder.}
\label{IEEE 13 system}
\end{figure}

The system topology and its corresponding node-cell-based graph representation are shown in Fig.~\ref{IEEE 13 system}~(a)~and~(b), respectively. As illustrated in Fig.~\ref{IEEE 13 system}~(a), in this system, there is a substation located in the source bus 650, with an additional Bus 670 that connects with Bus 632. 

\begin{figure}[!t]
\centering
\includegraphics[width=\columnwidth]{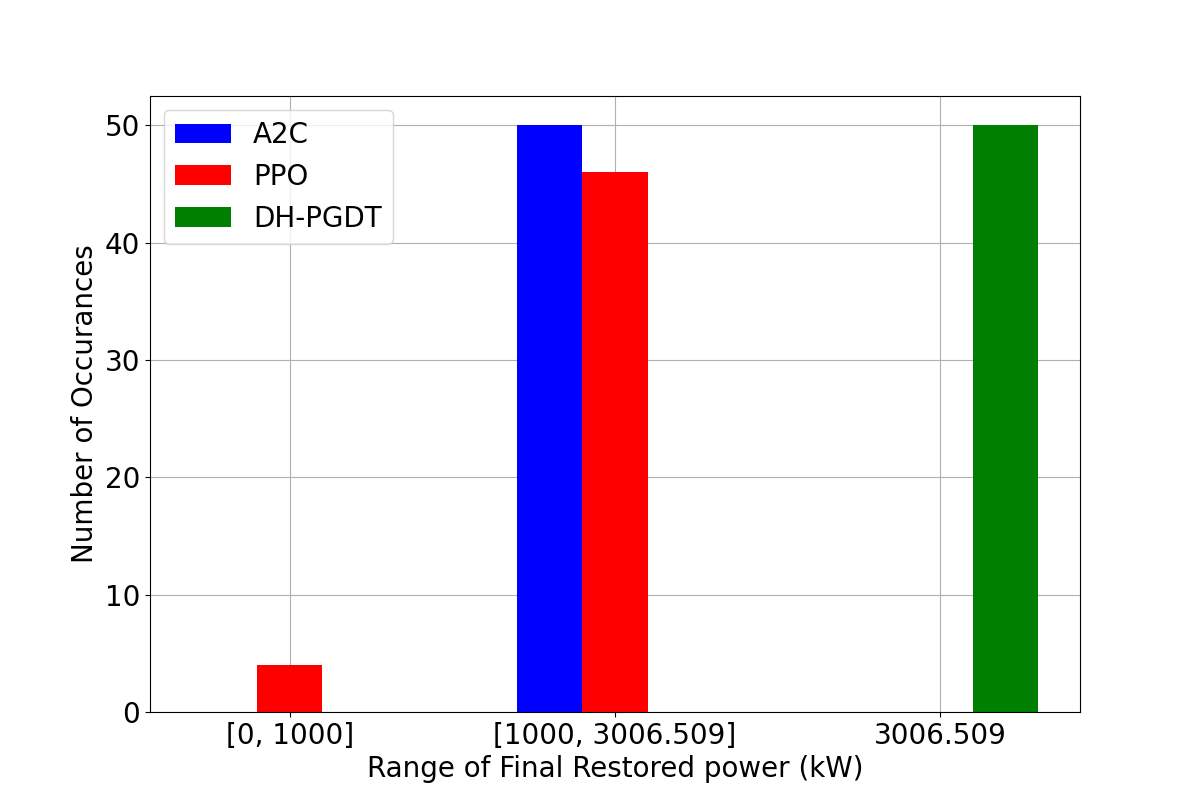}
\caption{Distribution of power restoration levels in the 50 independent trials using the three methods, respectively, in the modified 13-node test feeder.}
\label{IEEE 13 bar chart}
\end{figure}

In this test case, we used the energization path version of operational constraints to study the DSR problem. Due to its simple topological complexity, the ground-truth energization path in this case is determined as (Node Cell 0) → (Node Cell 1) → (Node Cell 3) → (Node Cell 4). And the corresponding final power demand should be $3006.509$~kW, which is considered the objective of the DSR operation. The evaluation results of the our method (we choose $q=2$ because the optimal solution only have 2 intermediate steps) and two DRL methods, A2C and PPO, across 50 independent inference trials are stated in Table~\ref{IEEE 13 results}. As shown in Table~\ref{IEEE 13 results}, we compare the performance of these three methods from four perspectives, including average return, standard deviation of returns (Std. Return), and number of expected optimal solutions (\# Opt. Sols.). Additionally, a bar chart in Fig.~\ref{IEEE 13 bar chart} shows more insights into the simulation results, which presents the distribution of power restoration levels in the 50 independent trials using the three methods, respectively. 

\begin{figure}[t]
\centering
\includegraphics[width=\columnwidth]{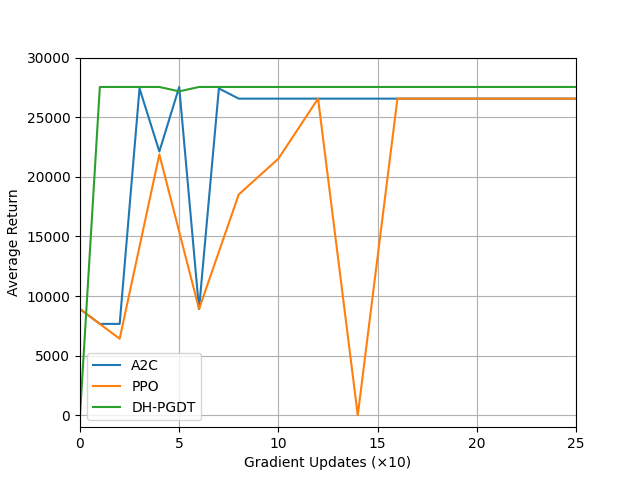}
\caption{Learning curves for the average returns of the first 3000 gradient updates using our DH-PGDT method and the two other benchmark DRL methods, PPO and A2C, for the DSR operations in the modified 13-node test feeder. The curve is updated per hundred gradient updates.}
\label{IEEE 13 learning curve}
\end{figure}

\begin{table}[!h]
\small
\setlength{\tabcolsep}{1mm}
\centering
\begin{tabular}{>{\centering\arraybackslash}m{1.2cm}|>{\centering\arraybackslash}m{1.5cm}|>{\centering\arraybackslash}m{1.5cm}|>{\centering\arraybackslash}m{0.5cm}}
\hline
\textbf{Method} & \textbf{Average Return} & \textbf{Std. Returns} & \textbf{\# Opt. Sols.} \\ \hline
\textbf{A2C} & 26557.891 & 0.000 & 0 \\ \hline
\textbf{PPO} & 25088.455 & 4787.505 & 1 \\ \hline
\textbf{DH-PGDT} & 27536.296 & 0.000 & 50 \\ \hline
\end{tabular}
\caption{Further performance comparison between our DH-PGDT method and two benchmark DRL methods for the DSR operation in the modified IEEE 13-node test feeder.}
\label{IEEE 13 results}
\end{table}

From Figs.~\ref{IEEE 13 learning curve} and~\ref{IEEE 13 bar chart} and Table~\ref{IEEE 13 results}, we can observe that, during the learning process, our method shows a faster convergence rate than the other two methods and achieves a higher convergence average return. Additionally, we can observe that A2C consistently converges to suboptimal solutions ranging between $1000$~kW and $3006.509$~kW across the 50 trials, while PPO shows a wider distribution of solutions but fails to achieve any optimal solutions in all 50 trials. Furthermore, the simulation results also show that the performance of the PPO method lies between that of A2C and our DH-PGDT methods.

\subsection{Modified IEEE 123-Node Test Feeder}
\begin{figure}[!h]
    \centering
    \includegraphics[width=\columnwidth]{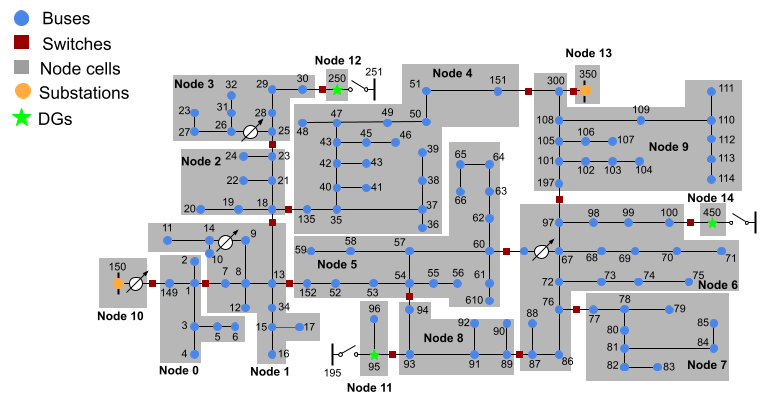}\\
    (a)\\
    \includegraphics[width=\columnwidth]{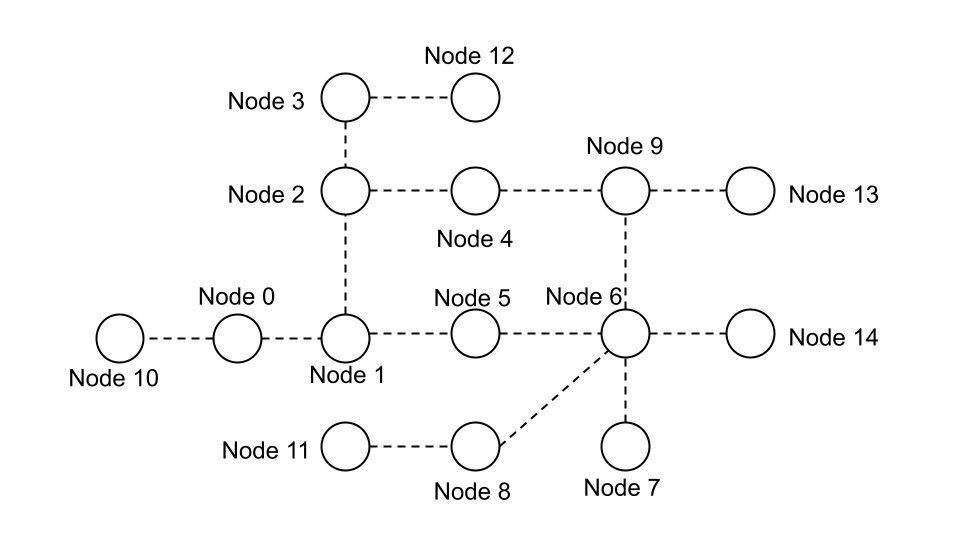}\\
    (b)
    \caption{(a): The physical topology of the modified IEEE 123-node test feeder; and (b): The node-cell-based graph representation of the modified IEEE 123-node test feeder.}
    \label{Topology of modified 123-node test feeder.}
\end{figure}

The system topology and the corresponding node-cell-based graph representation are shown in Fig.~\ref{Topology of modified 123-node test feeder.}~(a)~and~(b), respectively. As shown in Fig.~\ref{Topology of modified 123-node test feeder.}(a), in the modified 123-node test feeder, there are five energy sources in the system, two of which are substations located in Buses 150 and 350, and the three other sources are DGs located in Buses 95, 250, and 450. To evaluate the effectiveness of our DH-PGDT framework under complex operational scenarios, we introduce an additional operation constraint related to the energization capability of the DG at Bus 250. Specifically, this DG is limited to fully energizing only Node Cells 2 and 3. If it attempts to energize any other node cells beyond these, the DG automatically shuts down.

\begin{figure}[!t]
\centering
\includegraphics[width=0.9\columnwidth]{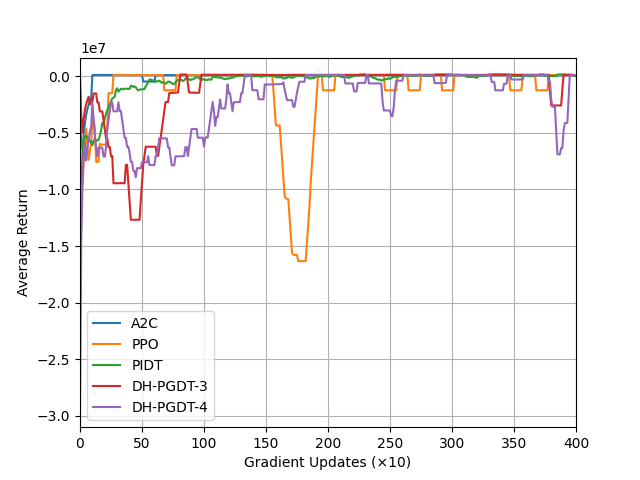}
\caption{Learning curve for the average return of the first 4000 gradient updates using our proposed DH-PGDT (DH-PGDT-$q$ denotes our DH-PGDT frameworks with hyperparameter $q$, representing the number of subgoals), PIDT, PPO, and A2C methods for the DSR operations in the modified 123-node test feeder. The curve is updated per ten gradient updates, and is smoothed using a moving average per 10 data points.}
\label{IEEE 123 learning curve}
\end{figure}

\begin{figure}[!t]
\centering
\includegraphics[width=0.9\columnwidth]{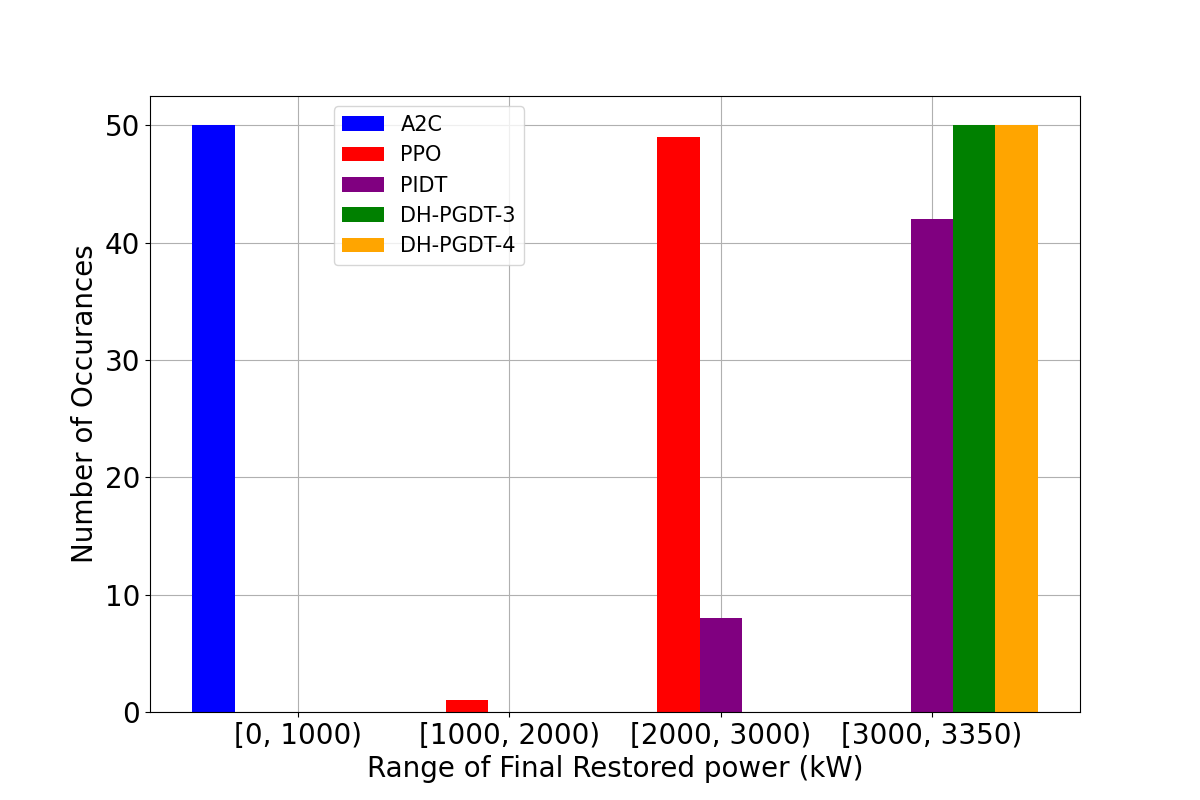}
\caption{Distribution of power restoration levels in the 50 independent trials using the four methods, respectively, in the modified 123-node test feeder.}
\label{IEEE 123 bar chart}
\end{figure}
The actual total power demand of the system is about $3350$~kW if all the loads are energized properly, which is considered as the objective of the DSR operation. The learning curves of the first 6000 gradient updates using the DH-PGDT, PIDT, A2C, and PPO methods are shown in Fig.~\ref{IEEE 123 learning curve}. As shown in Fig.~\ref{IEEE 123 learning curve}, for the modified IEEE 123-node test feeder with a complex and large-scale topology, our method has convergence speed comparable to the PIDT and two DRL methods. Fig.~\ref{IEEE 123 learning curve} also shows that the convergence speed of our method is influenced by the hyperparameter $q$, the predefined number of subgoals. As $q$ increases, the convergence becomes slower. Further evaluation results of these three methods across 50 independent trials in the inference stage are presented in Table~\ref{IEEE 123 results}. As shown in Table~\ref{IEEE 123 results}, our proposed DH-PGDT achieves optimal solutions in all 50 independent trials, outperforming PIDT, PPO and A2C.

\begin{table}[!h]
\small
\setlength{\tabcolsep}{1mm}
\begin{tabular}{>{\centering\arraybackslash}m{1.2cm}|>{\centering\arraybackslash}m{1.7cm}|>{\centering\arraybackslash}m{1.7cm}|>{\centering\arraybackslash}m{1.2cm}|>{\centering\arraybackslash}m{1.0cm}|>{\centering\arraybackslash}m{0.5cm}}
\hline
\textbf{Method} & \textbf{Average Return} & \textbf{Std. Return} & \textbf{APR (kW)} & \textbf{SDPR (kW)} & \textbf{\# Opt. Sols.} \\ \hline
\textbf{A2C} & 8696.511 & 0.000 & 486.619 & 0.000 & 0 \\ \hline
\textbf{PPO} & -163818.290 & 1878998.794 & 2097.300 & 83.751 & 0 \\ \hline
\textbf{PIDT} & 158850.797 & 9911.682 & 3228.657 & 206.415 & 32 \\ \hline
\textbf{DH-PGDT-3} & 164978.536 & 0.000 & 3349.696 & 0.000 & 50 \\ \hline
\textbf{DH-PGDT-4} & 164978.536 & 0.000 & 3349.696 & 0.000 & 50 \\ \hline
\end{tabular}
\caption{Further performance comparison between our proposed DH-PGDT, PIDT, A2C, and PPO methods for the DSR operation in the modified IEEE 123-node test feeder. Metrics include average return, standard deviation of returns (Std. Return), average power restoration (APR), standard deviation of power restorations (SDPR), and number of expected optimal solutions (\# Opt. Sols.).}
\label{IEEE 123 results}
\end{table}

Furthermore, Fig.~\ref{IEEE 123 bar chart} provides deeper insights into the simulation results shown in Table~\ref{IEEE 123 results} by illustrating the distribution of power restoration levels across 50 independent trials using the four methods. Our method consistently achieves optimal power restoration in all 50 trials. In contrast, the A2C method predominantly generates results in the range $1000$~kW to $2000$~kW. The output of the PPO method mainly falls within the range of $2000$~kW to $3000$~kW, occasionally dropping to the range of $1000$~kW to $2000$~kW. The output of the PIDT method mainly falls within the range of $3000$~kW to $3350$~kW, occasionally dropping to the range of $2000$~kW to $3000$~kW, which is better than both PPO and A2C, but worse than our currently proposed DH-PGDT method. These findings highlight the superior performance of our method compared to the PIDT method in our previous work and two emerging DRL methods.

\subsection{Modified Iowa 240-Node Test Feeder}
\begin{figure}[!h]
    \centering
    \includegraphics[width=\columnwidth]{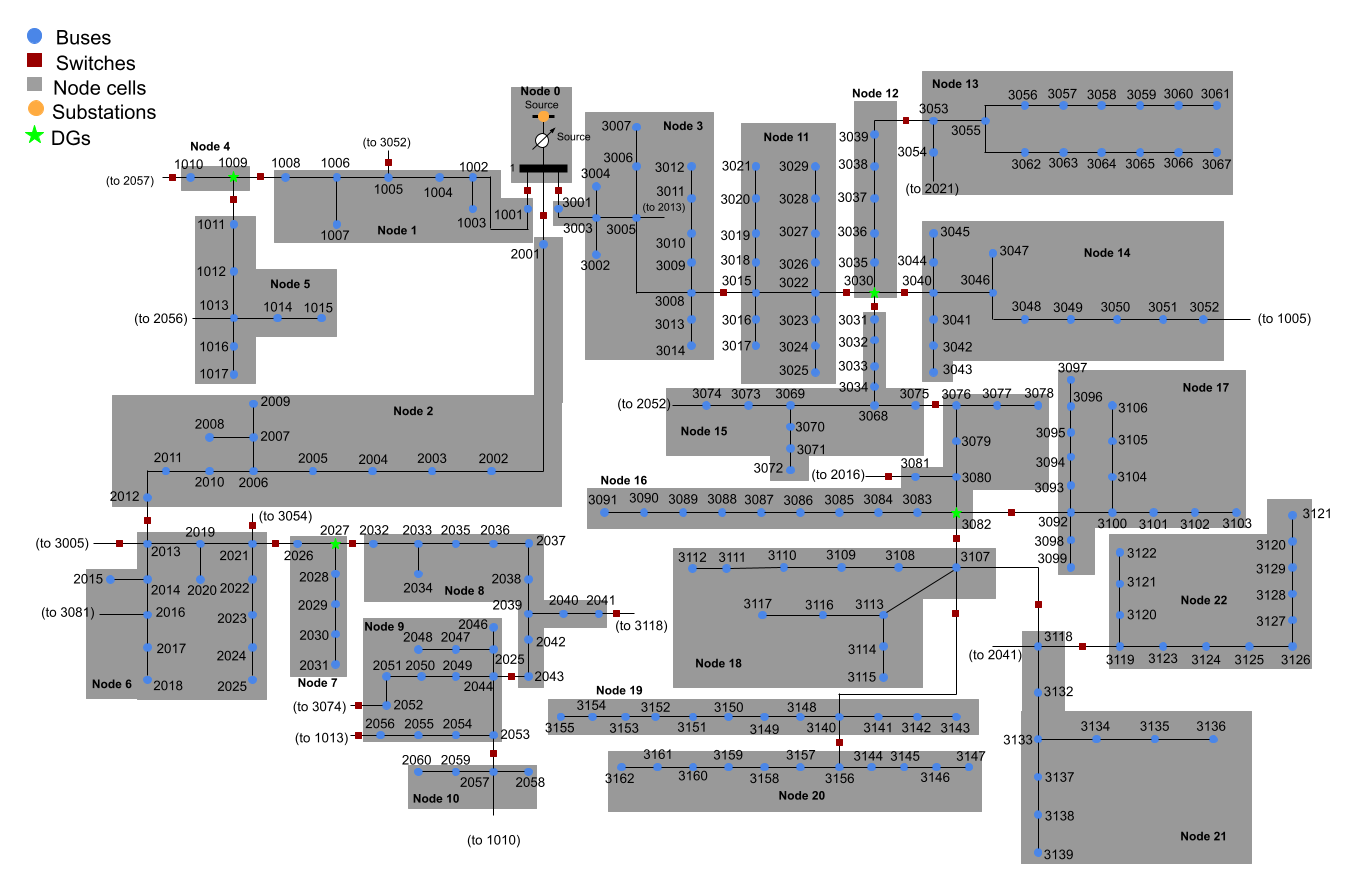}\\
    (a)\\
    \includegraphics[width=\columnwidth]{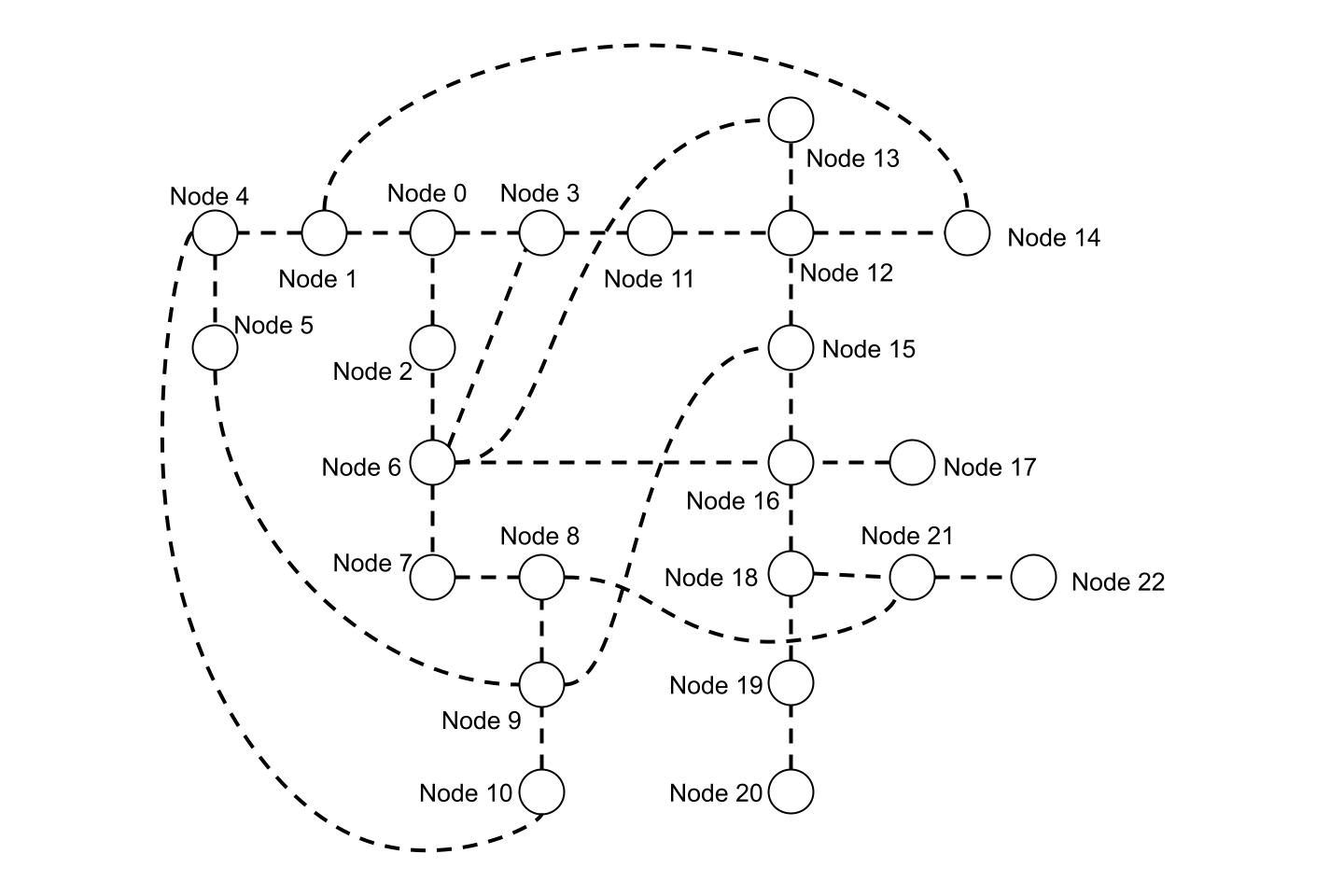}\\
    (b)
    \caption{(a): The physical topology of the modified Iowa 240-node test feeder; and (b): The node-cell-based graph representation of the modified Iowa 240-node test feeder.}
    \label{fig:Topology of modified Iowa 240-node test feeder.}
\end{figure}

The system topology and the corresponding node-cell-based graph representation are shown in Fig.~\ref{fig:Topology of modified Iowa 240-node test feeder.}~(a)~and~(b), respectively. As shown in Fig.~\ref{fig:Topology of modified Iowa 240-node test feeder.}~(a), the modified Iowa 240-node test feeder includes five energy sources in the system: one substation located at the source bus and four DGs located at Buses 1009, 3030, 2027, and 3082. To evaluate the zero-shot generalization capability of our DH-PGDT framework, we configure the system with dynamic loads whose active power varies over the time horizon. The model is trained solely on a static-load condition but tested directly on unseen dynamic load conditions without any fine-tuning.

\begin{figure}[!t]
\centering
\includegraphics[width=0.9\columnwidth]{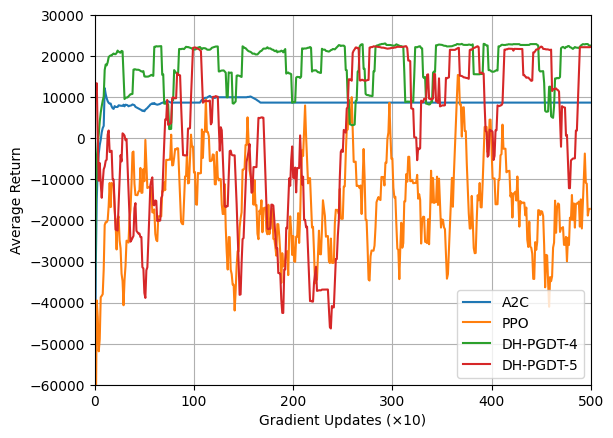}
\caption{Learning curve for the average return of the first 5000 gradient updates using our DH-PGDT method and two benchmark DRL methods, PPO and A2C, for the DSR operations in the modified Iowa 240-node test feeder. The curve is updated per ten gradient updates and is smoothed using moving average per 10 data points.}
\label{Iowa 240 learning curve}
\end{figure}

\begin{figure}[!t]
\centering
\includegraphics[width=0.9\columnwidth]{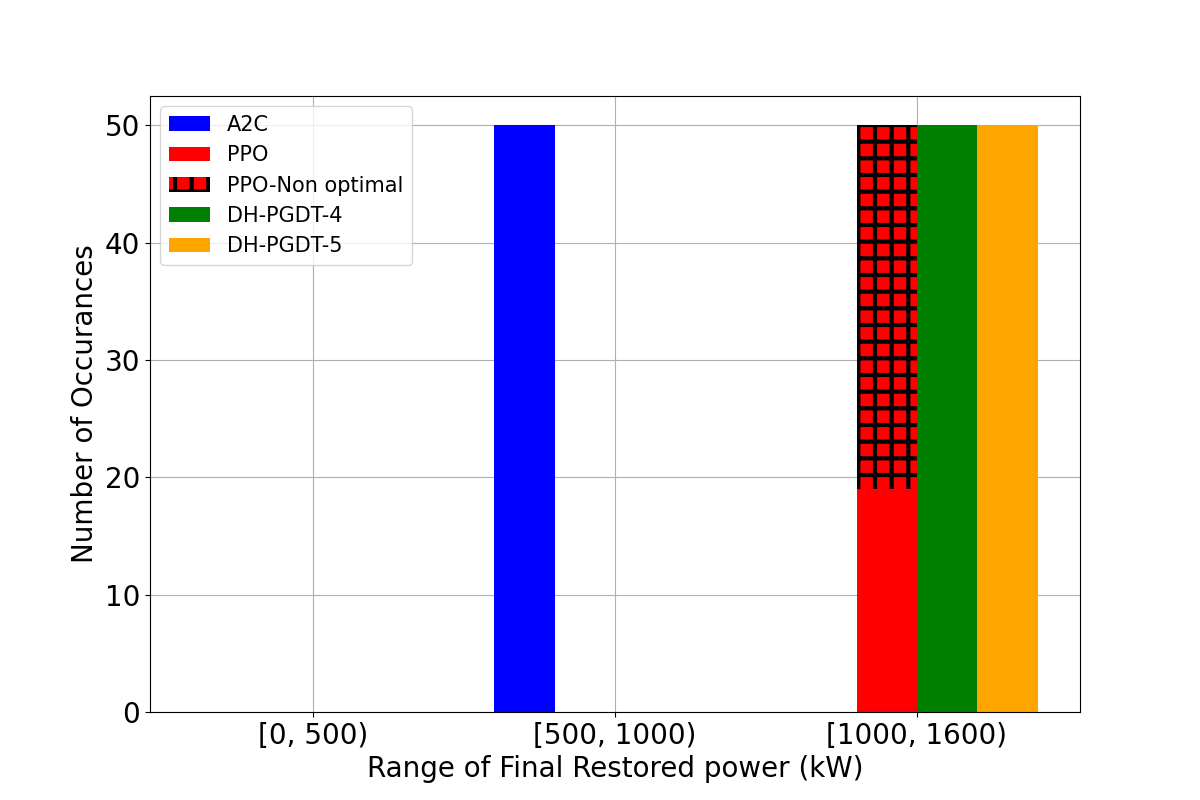}
\caption{Distribution of power restoration levels in the 50 independent trials using the three methods, respectively, in the modified Iowa 240-bus test feeder.}
\label{Iowa 240 bar chart}
\end{figure}

Unlike the 123-node test case, the dynamic nature of the 240-node test case makes it impossible to determine a single total power value without fixing the time horizon. Therefore, we set the time horizon to $T=20$, which results in a final total restored power of approximately $1565$~kW.

The learning curves of the first 5000 gradient updates using our DH-PGDT method and two benchmark DRL methods are shown in Fig.~\ref{Iowa 240 learning curve}. Additional evaluation results across 50 independent inference trials are presented in Table~\ref{Iowa 240 results}. As shown in Fig.~\ref{Iowa 240 learning curve} and Table~\ref{Iowa 240 results}, although our method achieves a convergence rate comparable to the two DRL methods, it consistently reaches optimal solutions in all 50 trials, which outperforms both the DRL methods. Furthermore, Fig.~\ref{Iowa 240 bar chart} provides deeper insights into the simulation results presented in Table~\ref{Iowa 240 results} by illustrating the distribution of power restoration levels across 50 independent trials using the three methods. It shows that our method achieves optimal power restoration in all 50 trials. In contrast, the A2C-based method predominantly produces solutions within the $500$--$1000$~kW range. While PPO mainly generates solutions in the $1000$--$1600$~kW range, it is worth noting that 31 of these solutions violate the ramp rate constraints, rendering them non-optimal. 

\begin{table}[t]
\small
\setlength{\tabcolsep}{1mm}
\begin{tabular}{>{\centering\arraybackslash}m{1.2cm}|>{\centering\arraybackslash}m{1.5cm}|>{\centering\arraybackslash}m{1.5cm}|>{\centering\arraybackslash}m{1.2cm}|>{\centering\arraybackslash}m{1.0cm}|>{\centering\arraybackslash}m{0.5cm}}
\hline
\textbf{Method} & \textbf{Average Return} & \textbf{Std. Return} & \textbf{APR (kW)} & \textbf{SDPR (kW)} & \textbf{\# Opt. Sols.} \\ \hline
\textbf{A2C} & 47414.036 & 1675.582 & 1077.232 & 0.000 & 0 \\ \hline
\textbf{PPO} & -18523.242 & 33103.943 & 1565.522 & 0.007 & 19 \\ \hline
\textbf{DH-PGDT-4} & 23203.563 & 0.000 & 1565.526 & 0.000 & 50 \\ \hline
\textbf{DH-PGDT-5} & 23330.984 & 0.000 & 1565.526 & 0.000 & 50 \\ \hline
\end{tabular}
\caption{Further performance comparison between our DH-PGDT method and benchmark DRL methods for the DSR operation in the modified Iowa 240-Node test feeder.}
\label{Iowa 240 results}
\end{table}

\begin{figure}[htbp]
\centering
\includegraphics[width=1.0\columnwidth]{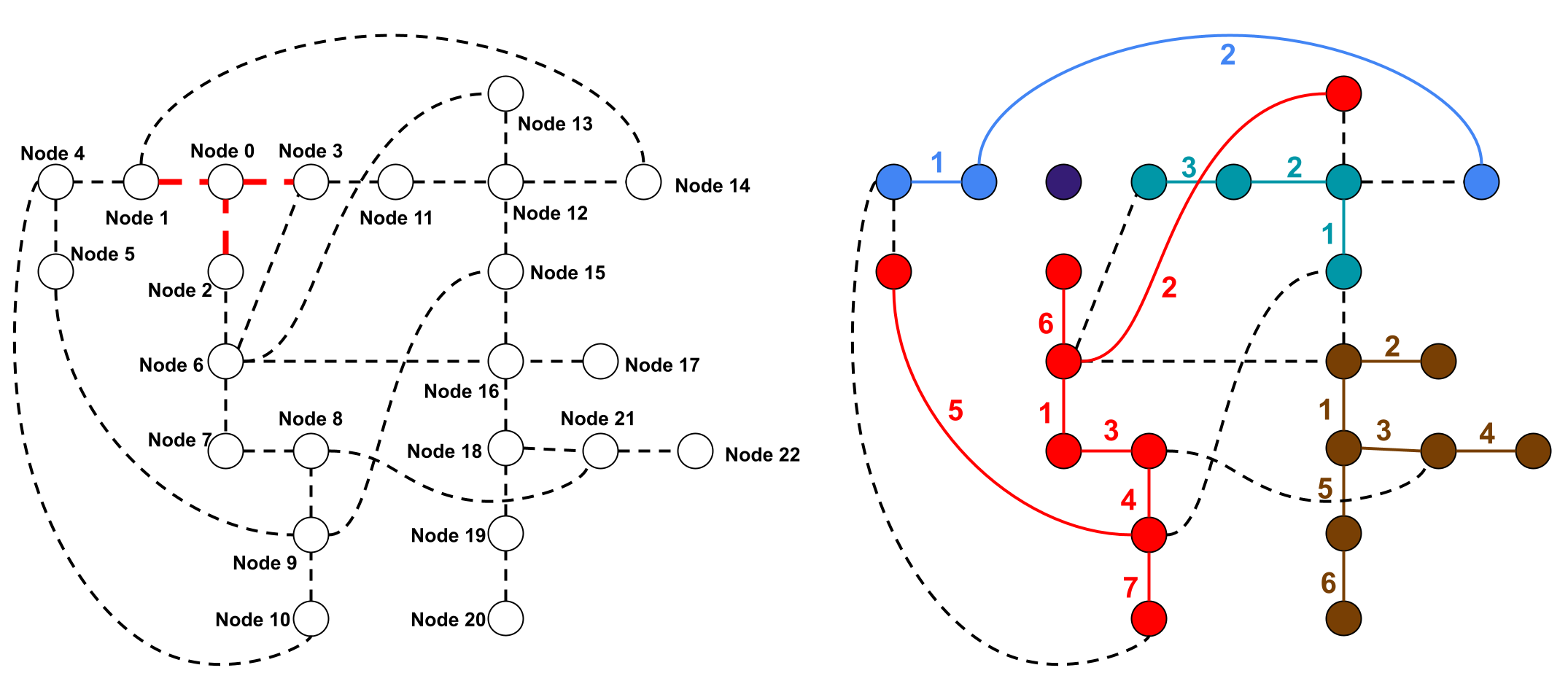}\\
(a)\qquad\qquad\qquad\qquad\qquad\qquad(b)
\caption{(a): Graph representation of Scenario 1, with red lines indicating switch contingencies; (b): Solution generated from DH-PGDT for Scenario 1, with numbers on the sides representing the step when the switch is closed.}
\label{fig:240 bus contingency scenario 1 with solution}
\end{figure}

\begin{figure}[!ht]
\centering
\includegraphics[width=1.0\columnwidth]{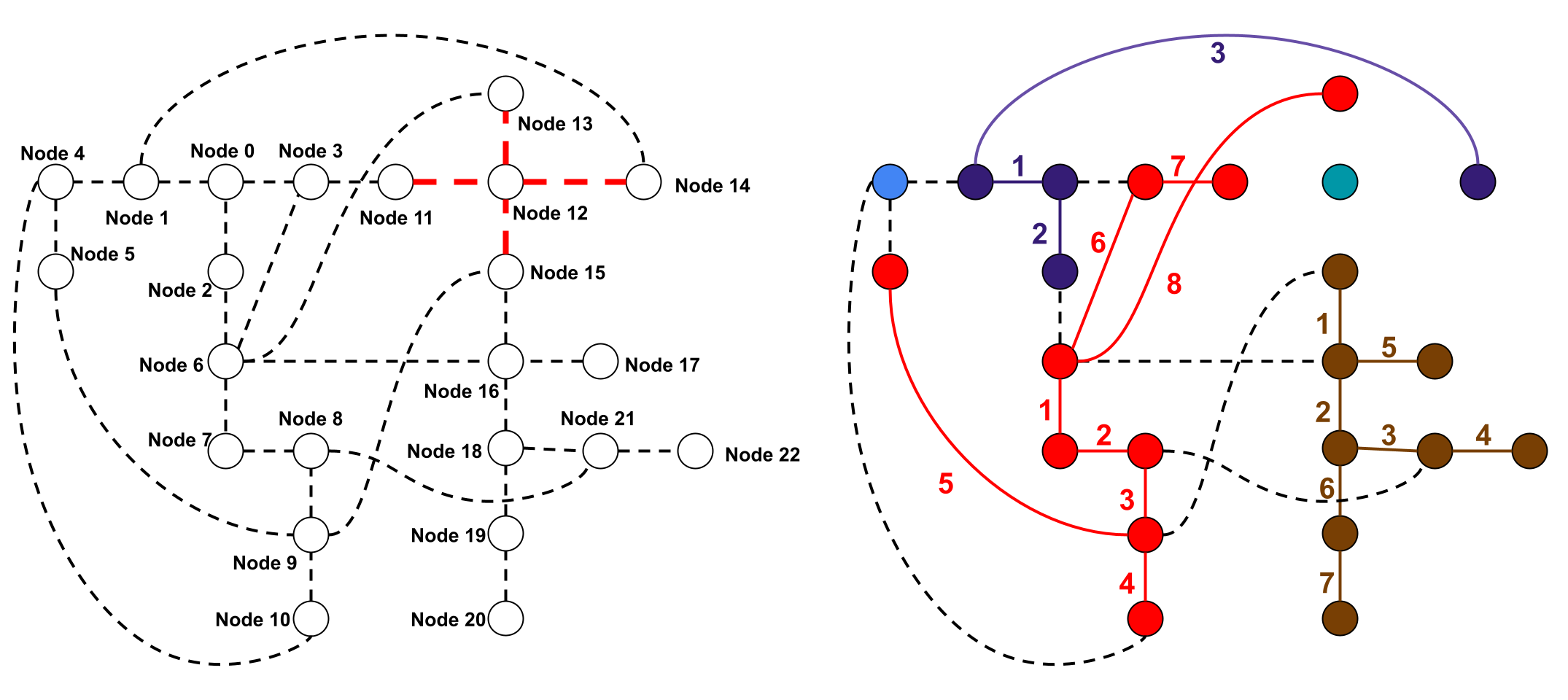}\\
(a)\qquad\qquad\qquad\qquad\qquad\qquad(b)
\caption{(a): Graph representation of Scenario 2, with red lines indicating the switch contingencies; (b): Solution generated from DH-PGDT for Scenario 2, with numbers on the sides representing the step when the switch is closed.}
\label{fig:240 bus contingency scenario 2 with solution}
\end{figure}

Moreover, to further evaluate the zero-shot generalization of our DH-PGDT framework, we test it in two additional scenarios involving unexpected switch contingencies during inference. In the first scenario, a switch contingency causes the power source at the source bus in Fig.~\ref{fig:Topology of modified Iowa 240-node test feeder.} to become isolated from the system. For clarity, we use the node-cell-based graph representation to illustrate the contingency as shown in Fig.~\ref{fig:240 bus contingency scenario 1 with solution}~(a), where Node Cell 0 is isolated from the rest of the graph. We would like to mention that the node in the graph corresponds to the node cells in the system topology in Fig.~\ref{fig:Topology of modified Iowa 240-node test feeder.}. The solution generated by our method to adapt to this zero-shot scenario is shown in Fig.~\ref{fig:240 bus contingency scenario 1 with solution}~(b), achieving a final restored power of $1565.526$~kW and a final return of $22743.817$. In the second scenario, a switch contingency causes the DG at Bus 3030 in Fig.~\ref{fig:Topology of modified Iowa 240-node test feeder.} to become isolated from the system. The contingency is illustrated using the node-cell-based graph representation as shown in Fig.~\ref{fig:240 bus contingency scenario 2 with solution}~(a), where Node Cell 12 is isolated from the rest of the graph. The solution generated by our method to adapt to this zero-shot scenario is shown in Fig.~\ref{fig:240 bus contingency scenario 2 with solution}~(b), achieving a final restored power of $1565.526$~kW and a final return of $22308.007$.
 
\section{Conclusions}
In this paper, we proposed DH-PGDT, an innovative framework for complex DSR tasks under uncertainty. By integrating a dual-head physics-informed causal transformer architecture with an operational constraint-aware graph reasoning module, the proposed DH-PGDT effectively enables scalable and reliable DSR even in zero-shot and few-shot scenarios. Extensive evaluations with different large-scale benchmark distribution systems demonstrate that DH-PGDT significantly improves restoration efficiency, constraint satisfaction, and generalization to unforeseen contingencies. While this framework is specifically designed for DSR in power systems, its underlying computing model is generalizable to a wide range of sequential decision-making tasks in engineering domains with complex physical and operational constraints. In ongoing work, we are evaluating the performance of DH-PGDT in even larger-scale DSR scenarios with diverse contingencies and extending the framework to other power system operational tasks.

\bibliography{aaai2026}

\end{document}